\documentclass[10pt,conference]{IEEEtran}
\IEEEoverridecommandlockouts
\usepackage{graphicx}
\usepackage{dcolumn}
\usepackage{bm}
\usepackage{amsthm}
\usepackage{amsmath}
\usepackage{amssymb}
\usepackage{color}
\usepackage{multirow}
\usepackage{algorithm}
\usepackage{algorithmicx}
\usepackage{algpseudocode}
\usepackage{subfigure}
\usepackage{hyperref}
\usepackage{wasysym}
\usepackage{xcolor}
\usepackage{booktabs}

\usepackage{float}
\usepackage{comment}
\usepackage{graphicx}

\def\BibTeX{{\rm B\kern-.05em{\sc i\kern-.025em b}\kern-.08em
    T\kern-.1667em\lower.7ex\hbox{E}\kern-.125emX}}
\begin{document}

\title{Quantum Graph Optimization Algorithm}


\makeatletter 
\newcommand{\linebreakand}{%
  \end{@IEEEauthorhalign}
  \hfill\mbox{}\par
  \mbox{}\hfill\begin{@IEEEauthorhalign}
}
\makeatother 

\author{
    \IEEEauthorblockN{Yuhan Huang}
    \IEEEauthorblockA{\textit{Department of Electronic and Computer Engineering}\\
    \textit{The Hong Kong University of Science and Technology}\\
    Hong Kong SAR, China}
    \and
    \IEEEauthorblockN{Ferris Prima Nugraha}
    \IEEEauthorblockA{\textit{Department of Electronic and Computer Engineering}\\
    \textit{The Hong Kong University of Science and Technology}\\
    Hong Kong SAR, China}
    \and

    \linebreakand
    
    \IEEEauthorblockN{Siyuan Jin}
    \IEEEauthorblockA{\textit{Department of Information Systems}\\
    \textit{The Hong Kong University of Science and Technology}\\
    Hong Kong SAR, China}
    \and
    \IEEEauthorblockN{Yichi Zhang}
    \IEEEauthorblockA{\textit{Department of Physics}\\
    \textit{The Hong Kong University of Science and Technology}\\
    Hong Kong SAR, China}
    \and

    \linebreakand
    
    \IEEEauthorblockN{Bei Zeng*}
    \IEEEauthorblockA{\textit{Department of Physics}\\
    \textit{The Hong Kong University of Science and Technology}\\
    Hong Kong SAR, China}
    \and
    \IEEEauthorblockN{Qiming Shao*}
    \IEEEauthorblockA{\textit{Department of Electronic and Computer Engineering}\\
    \textit{The Hong Kong University of Science and Technology}\\
    Hong Kong SAR, China}
}

\maketitle

\begin{abstract}

Quadratic unconstrained binary optimization~(QUBO) tasks are very important in chemistry, finance, job scheduling, and so on, which can be represented using graph structures, with the variables as nodes and the interaction between them as edges. Variational quantum algorithms, especially the Quantum Approximate Optimization Algorithm (QAOA) and its variants, present a promising way, potentially exceeding the capabilities of classical algorithms, for addressing QUBO tasks. However, the possibility of using message-passing machines, inspired by classical graph neural networks, to enhance the power and performance of these quantum algorithms for QUBO tasks was not investigated. This study introduces a novel variational quantum graph optimization algorithm that integrates the message-passing mechanism, which demonstrates significant improvements in performance for solving QUBO problems in terms of resource efficiency and solution precision, compared to QAOA, its variants, and other quantum graph neural networks. Furthermore, in terms of scalability on QUBO tasks, our algorithm shows superior performance compared to QAOA, presenting a substantial advancement in the field of quantum approximate optimization.

\end{abstract}


\section{Introduction}
Quantum computing has demonstrated tremendous potential in solving high-complexity problems in the NISQ (Noisy Intermediate-Scale Quantum) era~\cite{preskill2018quantum}, such as factorizing large numbers~\cite{shor1994algorithms}, solving linear systems of equations~\cite{harrow2009quantum} and others ~\cite{cai2021multi,arute2019quantum,madsen2022quantum,du2021learnability}. However, the noisy nature and limited number of reliable logical qubits in NISQ systems present significant obstacles in fully harnessing the power of quantum computing. To overcome these challenges, researchers have developed classical-quantum hybrid algorithms~\cite{cerezo2021variational} that combine classical optimization techniques with quantum computing techniques.

Variational quantum algorithms (VQAs), as one classical-quantum hybrid algorithm, involve a classical optimizer for quantum circuit parameters to minimize the objective function of certain problems. VQAs demonstrate great performance for both optimization and learning tasks: 1) optimization tasks: variational quantum eigensolver~\cite{kandala2017hardware} and quantum approximate optimization algorithms~\cite{farhi2014quantum} in quantum chemistry~\cite{kandala2017hardware,gokhale2020n} and finance~\cite{zoufal2023variational,brandhofer2023benchmarking}; 2) learning task: prediction and classification~\cite{hou2021universal,perez2020data}. In classical domain, VQAs could have applications in finance~\cite{kyriienko2022unsupervised,liu2022quantum}, natural language processing~\cite{lorenz2023qnlp}, and computer vision~\cite{jing2022rgb}. In the quantum domain, VQAs can solve problems such as symmetry-protected topological phase discrimination~\cite{li2024ensemble}, entanglement witness~\cite{scala2022quantum}, and state preparation~\cite{zhang2022quantum}.

Most of the existing VQAs algorithms are designed to process Euclidean distance data, which rely on standard straight-line metrics. However, there is a growing interest in developing VQAs that can process non-Euclidean data constrained to graphs. In graph-structured data, connections between nodes go beyond simple linear distances. The data is represented as a graph, where each node has its own set of features, and the connections between pairs of nodes are encoded as edge features. This representation has proven to be essential in a wide range of sophisticated applications, such as social network analysis~\cite{hamilton2017inductive}, protein structures~\cite{strokach2020fast}, traffic networks~\cite{jiang2022graph}, and portfolio assignment~\cite{schuetz2022combinatorial}.

Quantum graph neural networks (QGNNs)~\cite{verdon2019quantum,ai2022decompositional,mernyei2022equivariant,zheng2021quantum,farhi2014quantum} are novel VQAs designed to utilize quantum computers to process graph-structured data. QGNNs have been applied in several scenarios, such as learning quantum Hamiltonian dynamics~\cite{verdon2019quantum}, graph isomorphic classification~\cite{verdon2019quantum}, learning-based discrete traveling salesman problem~\cite{skolik2023equivariant}, and graph classification~\cite{schatzki2024theoretical}. These applications demonstrate the potential of QGNNs to handle graph-based tasks. For quadratic unconstrained binary optimization~(QUBO) problems that can be represented as graphs with nodes and edges, the quantum approximate optimization algorithm~(QAOA) cannot take full advantage of the intrinsic properties of the graph structure. Therefore, how to design a quantum graph optimization algorithm specifically for the QUBO problem to enhance the capability of quantum computing is still a problem worth exploring.

In this work, we leverage the geometric properties of graphs to facilitate message-passing, a mechanism where nodes exchange and refine information with neighbors through edges. Inspired by traditional graph convolutional networks~\cite{wu2019simplifying}, we've developed a quantum circuit in Hilbert space tailored for processing graph data. Our methodology enhances the power of quantum computing and provides deeper insights into the intrinsic behaviors of message propagation across the graph. Initially, the information pertaining to the edges and vertices of the graph is encoded into a Hamiltonian model. This model facilitates the propagation of information throughout the graph's topology. Subsequently, we construct trainable block operations that are capable of updating and aggregating information, thereby updating the node's information. Following the quantum measurement process, we use a classical optimizer to update the parameters of trainable feature updating and aggregation operations.

The methodology section elaborates on the conversion process from the classical QUBO problem to the Ising model. We describe two methods for embedding classical information into quantum systems. We detail our approach for transferring the classical graph neural network, denoted as G(g,x), into a variational quantum graph algorithm framework. This includes methods for encoding the adjacency matrix (g) and node features (x) into a quantum system. Additionally, we outline the optimization techniques used. 

In the numerical results section, we evaluate the effectiveness of our approach for the portfolio assignment and minimum vertex cover problem. For the QUBO problem, we compare our approach with the commonly used graph-based optimization algorithms, QAOA~\cite{farhi2014quantum}, and QAOA variants~\cite{hadfield2019quantum,wang2020x}, to demonstrate our method's scalability and performance. 

\section{Preliminaries}

\subsection{Variational Quantum Algorithm}
VQAs are a class of quantum algorithms that combine classical and quantum computing resources to find approximate solutions to complex problems. VQAs are particularly useful for tasks that are computationally challenging for classical computers, such as simulating quantum systems or solving large-scale linear algebra problems.

VQAs work by using a parameterized quantum circuit to represent a trial solution to the problem at hand. The parameters of this circuit are then optimized using a classical optimization algorithm to minimize some cost functions. This allows VQAs to find approximate solutions to complex problems without requiring the full power of a fault-tolerant quantum computer.

Several different categories of VQAs have been developed, each targeting different types of problems:

\textbf{Variational quantum eigensolvers (VQEs)} are a type of VQA that can be used to find the ground state energy of a quantum many-body system~\cite{liu2024variational} or a chemical molecule~\cite{kandala2017hardware, gokhale2020n}. This is achieved by encoding the Hamiltonian of the system into the cost function and then optimizing the quantum circuit to minimize this cost function.

\textbf{Variational quantum classifiers (VQCs)} are a type of VQA that can be used to solve classical classification problems, such as MNIST~\cite{yano2021efficient}, community detection~\cite{nembrini2022towards}, fraud detection~\cite{innan2024financial} or quantum classification problems, such as entanglement witness~\cite{scala2022quantum} or the phase of matter predicting~\cite{liu2023model}. In this case, the quantum circuit is used to encode the features of the classification problem, and the optimization process is used to learn an optimal classification boundary.

\textbf{Quantum graph neural networks} are a type of VQA that can be used to solve problems defined on graphs, such as graph isomorphism or maximum cut problems. QGNNs use a quantum circuit to represent the graph and then optimize the parameters of this circuit to solve the problem.

\textbf{Quantum approximate optimization algorithm} is a typical quantum-classical hybrid algorithm to solve combinatorial optimization problems. QAOA uses a parameterized quantum circuit including the cost Hamiltonian $\text{H}_\text{C}$ and mixer Hamiltonian $\text{H}_\text{B}$ to prepare a quantum state that encodes the solution to the optimization problem. 

The $\text{H}_\text{C}$ in QAOA is typically expressed using the Ising model, which is a physical model of interacting spins on a lattice. The Ising model is used to represent the optimization problem as a collection of interacting binary variables, with the cost Hamiltonian given by:

\begin{align}
   \text{H}_\text{C} = \sum_{i<j} w_{ij}Z_iZ_j + \sum_i h_i Z_i
\end{align}
where $w_{ij}$ and $h_i$ are the coefficients, and $Z_i$ and $Z_j$ are Pauli Z operators acting on the $i$-th and $j$-th qubit. The first term in the cost Hamiltonian represents the interaction between pairs of particles, while the second term represents the bias towards certain particles.

The $\text{H}_\text{B}$ in QAOA is typically expressed as the Pauli X operators acting on each qubit and is given by:
\begin{align}
   \text{H}_\text{B} = \sum_{i=1}^{n} X_i
\end{align}
where $n$ is the number of qubits in the quantum circuit. However, other simple Hamiltonian can also be used for the mixer.

In QAOA, the initial state of the quantum circuit is typically chosen to be the ground state of the mixer, such as the $|+\rangle$ state of the mixer Hamiltonian $\text{H}_\text{B}$. This is because QAOA can be viewed as a type of adiabatic evolution, where the quantum circuit slowly evolved from the initial state to the final state that encodes the solution to the optimization problem. By starting in the ground state of a mixer, the quantum circuit remains in the lowest energy of the Hamiltonian throughout the evolution, which can find the optimal solution for cost $\text{H}_\text{C}$ problems.

\subsection{Quadratic Unconstrained Binary Optimization} 

Quadratic unconstrained binary optimization is a mathematical optimization problem that involves minimizing a quadratic objective function $(x^TAx)$ by finding the optimal binary values ($x=0$ or $x=1$) for a set of variables.

For solving the QUBO problem, there are both learning-based and non-learning-based algorithms in quantum. In Tab.~\ref{tab:gnn}, several algorithms for the QUBO problem are listed. One graph-based learning quantum algorithm is EQC~\cite{skolik2023equivariant}, which exploits the symmetry in graph data and designs an equivariance-based quantum graph neural network. On the other hand, our proposed algorithm takes a graph-based non-learning approach, called the quantum graph optimization algorithm, to achieve higher accuracy in solving the QUBO problem. The non-graph-based learning algorithm is T-QAOA~\cite{mahroo2023learning}, which utilizes an LSTM to train the QAOA ansatz for solving the QUBO problem. As for QAOA~\cite{farhi2014quantum} and its variants ~\cite{wang2020x,hadfield2019quantum}, it falls into the non-graph-based non-learning model category. It is important to note that although QAOA and its variant involve graph-related information in $H_C$, their original papers did not specifically design them with graph structures in mind, but rather they are based on the adiabatic process and did not fully consider utilizing graph logic to update the results.

\begin{table}[htbp]
  \centering
  \caption{Quantum algorithms for solving QUBO problem}
  \renewcommand{\arraystretch}{2}
  \begin{tabular}{|c|c|c|}
    \hline
    & Graph & Non-Graph \\
    \hline
    Learning & EQC~\cite{skolik2023equivariant} &  T-QAOA~\cite{mahroo2023learning} \\
    \hline
    \multirow{2}{*}{Non-Learning} & \multirow{2}{*}{This work} & QAOA~\cite{farhi2014quantum} \\
     & & XYQAOA~\cite{wang2020x}, BitflipQAOA~\cite{hadfield2019quantum} \\
     \hline
  \end{tabular}
  \label{tab:gnn}
\end{table}

\section{Quantum Graph Optimization Algorithm}
The Quantum Graph Optimization Algorithm (QGOA) is a variational quantum approximate optimization algorithm designed to address graph-related problems, which encodes the topology information of the graph-related problem into the Hamiltonian. QGOA has four primary stages: converting the graph problem into the Ising model, encoding the graph's topology into the quantum Hamiltonian, designing the quantum circuit, and optimizing the solution using a variational approach.

\begin{figure}[htp]
    \centering
    \includegraphics[width=1\columnwidth]{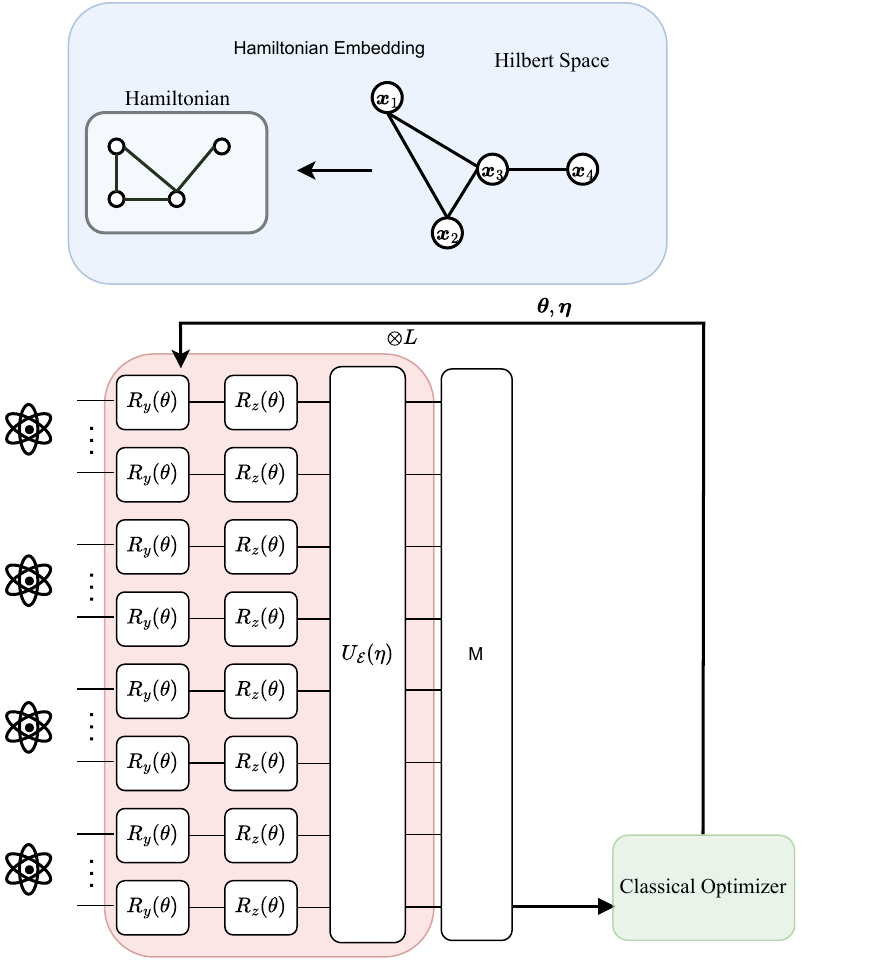}
    \caption{The blue block describes Hamiltonian encoding. The graph quantum circuit design consists of three components. 1) the trainable block with two layers of single-qubit gates, and one layer of the parameterized Hamiltonian evolution governed by the graph topology. 2) measurement operation. 3) the optimization step requires a classical optimizer.}
    \label{fig:GQAD}
\end{figure}

\subsection{Converting the Graph Problem into the Ising Model}

The first step of QGOA involves converting the graph problem into an Ising model representation. The graph-related problems discussed in this work involve the QUBO problem, which can be formulated as an optimization problem with binary variables and a quadratic objective function. Specifically, the QUBO problem can be expressed as:

\begin{align}
\ell(x)= \arg \min_x \sum_i a_{ii} x_i^2+\sum_{ij} a_{ij} x_ix_j+\sum_i b_i x_i,
\end{align}
where the coefficients $a$ and $b$ represent the weights of each term. To solve this problem using quantum computing, we can convert it into the Ising model. We establish a relationship between the binary variables and the Pauli Z operators by using the mapping $Z_i = 2x_i - 1$. This mapping associates a binary variable $x_i$ with a Pauli Z operator $Z_i$. By applying this mapping, the objective function of the binary quadratic problem can be represented as:

\begin{align}
M= \sum_i (a_{ii}+b_i) (Z_i+1)/2+\sum_{ij} a_{ij} (Z_i+1)(Z_j+1)/4.
\label{eq. M}
\end{align}
where $Z_i$ represents the energy contribution from the individual qubits in the system and $Z_iZ_j$ represents the energy contribution from the interactions between pairs of qubits.

\subsection{Quantum Embedding}

Next, we explain how to encode the topological structure of the graph into the quantum Hamiltonian, shown in the blue block of Fig.~\ref{fig:GQAD}. For general~\cite{verdon2019quantum} graph-related problems, we encode the topological structure of the graph $G=(\mathcal{V}, \mathcal{E})$ (where $\mathcal{V}$ represents the set of vertices and $\mathcal{E}$ represents the set of edges) into the quantum Hamiltonian, to effectively leverage the graph information within the quantum computing framework. The general form of the graph Hamiltonian can be defined as:

\begin{align}
H(\theta) = \sum_{i,j \in E} \sum_{r \in I_{ij}} W_{rij} O_i^{r} \otimes P_j^{r} + \sum_{v \in V} \sum_{r \in J_v} B_{rv} R_v^{r},
\end{align}
where $W_{rij}$ is a weight associated with the edge between vertices $i$ and $j$, and $O_i^{r}$ and $P_j^{qr}$ are operators acting on the $i$-th and $j$-th qubits, respectively. The $B_{qrv}$ is a weight associated with vertex $v$, and $R_v^{r}$ is an operator acting on the $v$-th qubit. 

Specifically, we can map the adjacency matrix of the graph onto the $\sum_{i,j \in E} \sum_{r \in I_{ij}} W_{rij} O_i^{r} \otimes P_j^{r}$ terms. Each adjacency matrix element $E_{ij}$ represents the presence or absence of an edge between vertices $i$ and $j$, with a value of 1 or $w_{ij}$ indicating the presence of an edge, and 0 indicating the absence of an edge. Additionally, we can encode the vertex attribute information into the $\sum_{v \in V} \sum_{r \in J_v} B_{rv} R_v^{r}$ terms. The detailed design of the Hamiltonian for the QGOA will be illustrated in Section \ref{sec:design QGOA}.

\subsection{Designing Quantum Graph Circuit}\label{sec:design QGOA}
The third stage of QGOA involves designing a quantum circuit. The variational quantum circuit, combines non-parameterized and parameterized quantum gates, such as X, Y, CNOT, Rx, and CRx, to efficiently handle both classical and quantum computing tasks.

Existing VQAs include hardware efficient ansatz~\cite{kandala2017hardware}, XX-ZX ansatz~\cite{farhi2018classification}, and traditional quantum ansatz~\cite{kyriienko2021solving}. These works introduce VQA architecture that includes a feature encoding ansatz and parameterized quantum blocks. However, for the graph-related problems $G=(\mathcal{V}, \mathcal{E})$, capturing and exchanging the geometric information of the graph, the traditional VQAs are not enough. Classical Graph neural networks consider using feature updating and aggregation operations to update and pass the information of graph data. In this work, the feature represents the solution obtained from QGOA and encapsulates the information encoded within the variational quantum circuit, which is crucial for addressing the graph-related problem. To accomplish this, we construct the VQA, which includes updating $U(\theta)$ and aggregation $U_\epsilon(\eta)$ trainable block, as shown in Fig.~\ref{fig:GQAD}.

Feature updating process in QGOA updates the features of each node using two kinds of single-qubit gates, $R_y(\theta)=exp(-i\theta/2\sigma_y)$ and $R_z(\theta)=exp(-i\theta/2\sigma_z)$, which together form a universal single-qubit gate. This is done through a trainable function that combines the two gates, and the result is used to update the features of each node.

Feature aggregation function in QGOA is also critical, as it facilitates effective message-passing and captures the interactions within the graph-related problem. In classical graph neural networks, aggregation is typically straightforward and easily interpretable, but in quantum machine learning, it can be challenging to construct. QGOA proposes an H-Model to represent the connections within a graph and achieve quantum aggregation. The aggregation function is defined as follows:

\begin{align} 
\text{U}_{H}(\eta)|\boldsymbol{x}\rangle=\sum_i \sum_{|n_i|} f_{\eta j}(n_{ij})+f_{\eta,nn_i}(nn_i,:)|i\rangle, 
\label{eq:agg}
\end{align}
where the $n_i$ represents the neighbors of the $i$-th vertex and the $nn_i$ represents the non-neighbors of the $i$-th vertex. This equation directly aggregates nearby information with the single-parameterized term, while global information appears in the multi-parameterized terms.

In our approach, we draw inspiration from the classical graph neural network to encode a graph's adjacency matrix (g) into the Hilbert space. The off-diagonal elements of the matrix are represented by the quantum states $E_{ij}(|1\rangle\langle 0|_i \otimes |0\rangle\langle 1|_j + |0\rangle\langle 1|_i \otimes |1\rangle\langle 0|_j)$, corresponding to the edges between different nodes $i$ and $j$. Meanwhile, the diagonal elements are represented as $E_{ii}|1\rangle\langle 1|_i$. The encoded Hamiltonian is represented as follows:

\begin{align}
\text{H}  &= \sum_{(i,j) \in \mathbb{E}} \text{E}_{ij}(|1\rangle\langle 0|_i\otimes |0\rangle\langle 1|_j+|1\rangle\langle 0|_j\otimes |0\rangle\langle 1|_i)\notag\\
&+\sum_{i \in \mathbb{V}} \text{E}_{ii}|1\rangle\langle 1|_i,
\label{H}
\end{align}
where $E_{ij}$ is the entry in the adjacency matrix corresponding to the edge $(i,j)$, $\mathbb{E}$ is the set of all edges in the graph, and the $E_{ii}$ is the covariance of the node $i$. This Hamiltonian is equal to the Model $H_{T}$: 

\begin{align}
    \text{H}_{T}=\sum_{(i,j) \in \mathbb{E}} \text{E}_{ij} (\sigma_i^x \sigma_j^x + \sigma_i^y \sigma_j^y) -\sum_{i \in \mathbb{V}} \text{E}_{ii} \sigma_i^z
    \label{H_T}
\end{align}
where $\sigma_i^x$, $\sigma_i^y$, and $\sigma_i^z$ are the Pauli-X, Pauli-Y, and  Pauli-Z acting on the $i$-th vertex. the equivalence proof of Eq.\ref{H} and Eq.\ref{H_T} is presented in Appendix.~\ref{appendix:a}.

Due to the non-commutativity of the interactions in the $H_{T}$ model, $[{\text{H}_{T}}_{i,j}, {\text{H}_{T}}_{k,v}] \neq 0$. This means that the order in which the topological structure is embedded into the Hamiltonian can affect the final result of the algorithm since the non-commutativity of the interactions implies that the operations corresponding to different edges or vertices do not commute. This leads to different outcomes for different embedding orders. Therefore, while the $\text{H}_{T}$ model is a useful tool for studying quantum many-body systems, it may not be suitable for certain applications that require strict geometric equivariance in the Hamiltonian embedding. Based on the theory of geometric learning, we designed a first-order $\text{H}_{T}$ model as the geometric equivariance embedding model, equivalent to Eq.~\ref{H_T}. The advantage of this approach is that the terms in  $\sum_{(i,j)\in \mathcal{E}} \text{E}_{ij} \text{H}_{ij}^k$ mutually commute.

\begin{align}
    \text{H}_{T^{1st}}=\sum_k \sum_{(i,j)\in \mathcal{E}} \text{E}_{ij} \text{H}_{ij}^k-\sum_{i \in \mathbb{V}} \text{E}_{ii} \sigma_i^z,
\end{align}
where $k\in \{1,2\}$. When $k=1$, $\text{H}=\text{XX}$, and when $k=2$, $\text{H}=\text{YY}$. Not only does it encode the topological structure of the graph, but it also guarantees the equivariance of topology quantum embedding. 

The equivalence of the $\text{H}_{T^{1st}}$ and the aggregation function, as defined earlier, is non-trivial. We provide a detailed analysis of this equivalence in Appendix.~\ref{app: quantum aggregation}, where we demonstrate the correspondence between the quantum encoding of the graph's adjacency matrix and the theoretical aggregation model.

\subsection{Optimizing the Quantum Circuit}
Finally, to optimize the binary quadratic problem encoded as the Ising model in Eq.~\ref{eq. M}, we seek to minimize its expectation value, which is the objective function. We use the ADAM optimizer with $T$ iterations as the classical optimizer, and the optimization problem can be formulated as:

\begin{align}
\ell(\theta,\eta) =\arg\min_{\theta,\eta} \langle 0| U^\dagger(\theta,\eta) M U(\theta,\eta) |0\rangle,
\end{align}
where $U(\theta,\eta)$ is the quantum graph circuit with trainable parameters $\theta$ and $\eta$, and $|0\rangle$ is the initial state of the quantum circuit.

\section{Simulation Results}

In this section, we employ the QGOA to address the portfolio optimization problem and minimum vertex cover optimization problem.

\subsection{Portfolio Optimization}

\subsubsection{Portfolio Optimization Problem} 
Portfolio optimization is to find the optimal combination of assets to achieve a desired level of return while minimizing risk. This problem can be formulated as a mathematical optimization problem with binary decision variables, where each variable represents whether an asset is included or not included in the portfolio. The objective function for portfolio optimization is typically expressed as a combination of the expected return and the portfolio risk, which can be written as:

\begin{align}
   \ell(x) = \arg\min_{\boldsymbol{x}} \lambda \boldsymbol{x}^T V \boldsymbol{x} - (1 - \lambda) \mu^T \boldsymbol{x} ,
   \label{obj}
\end{align}
where $\boldsymbol{x}$ is a vector of binary decision variables, with $x_i$ representing whether asset $i$ is included in the portfolio. The covariance matrix $V$ captures the pairwise relationships between each asset's returns, and the diagonal elements represent the asset variances. The vector $\mu$ represents the expected returns for each asset, and $\lambda$ is a parameter that controls the trade-off between portfolio risk and return. The first term, $\lambda \boldsymbol{x}^T V \boldsymbol{x}$, represents the portfolio risk, which is minimized subject to the constraints. The second term, $(1 - \lambda) \mu^T \boldsymbol{x}$, represents the portfolio return, which is maximized subject to the constraints.

\subsubsection{QGOA for Portfolio Optimization} 
For the nine-qubit portfolio optimization problem, the graph quantum algorithm design consists of two blocks: two layers of single rotation gates and a layer of first-order $\text{H}_{T^{1st}}$-model. Next, the algorithm uses an observable M that is related to the problem. The objective function can be expanded as shown in Eq.~\ref{obj}:

\begin{align}
   \ell(x) = \arg\min_{\boldsymbol{x}} \sum_i \lambda V_{ii} x_i^2+\sum_{ij} \lambda V_{ij} x_i x_j-\sum_i (1-\lambda) \mu_i x_i,
   \label{eq finance}
\end{align}

To embed this problem into the observable, we define M as:

\begin{align}
   M&=\sum_i(\lambda V_{ii}-(1-\lambda) \mu_i)((Z_i+1)/2)\notag\\
   ~~~~~&+\sum_{ij}\lambda V_{ij}((Z_i+1)(Z_j+1)/4),
\end{align}
where Z represents the Pauli-Z operator; after measuring using observable M, we obtain classical information, which we then use to optimize the parameters $\theta$ and $\eta$ of the graph quantum circuit using the Adam optimizer.

\begin{figure*}[htbp]
    \centering
    \subfigure[]{
    \begin{minipage}[t]{0.35\textwidth}
    \centering
    \includegraphics[width=1\textwidth]{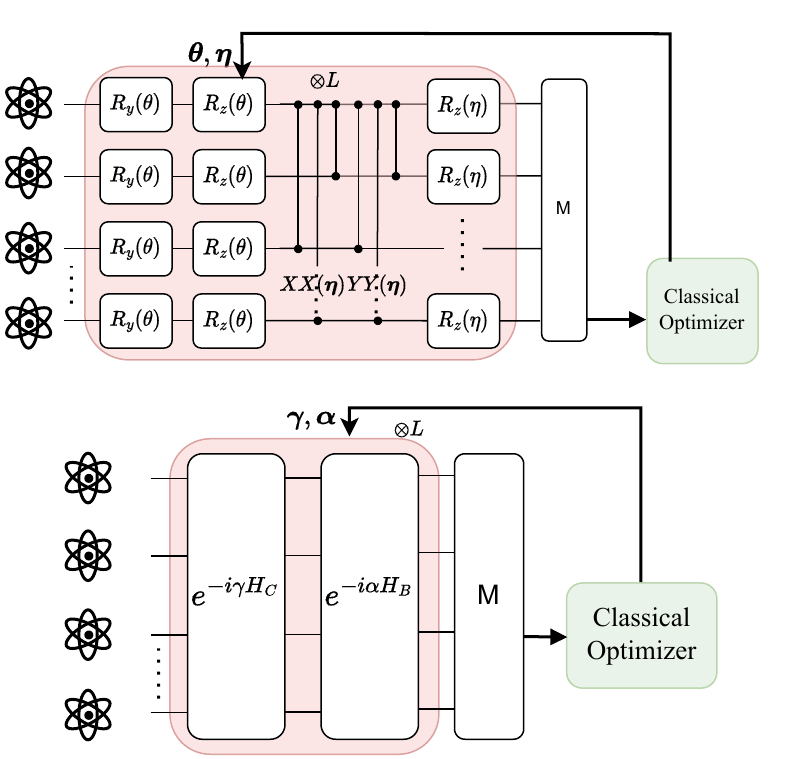}
    \end{minipage}\label{fig: finance}}
    \subfigure[]{
    \begin{minipage}[t]{0.6\textwidth}
    \centering
    \includegraphics[width=1\textwidth]{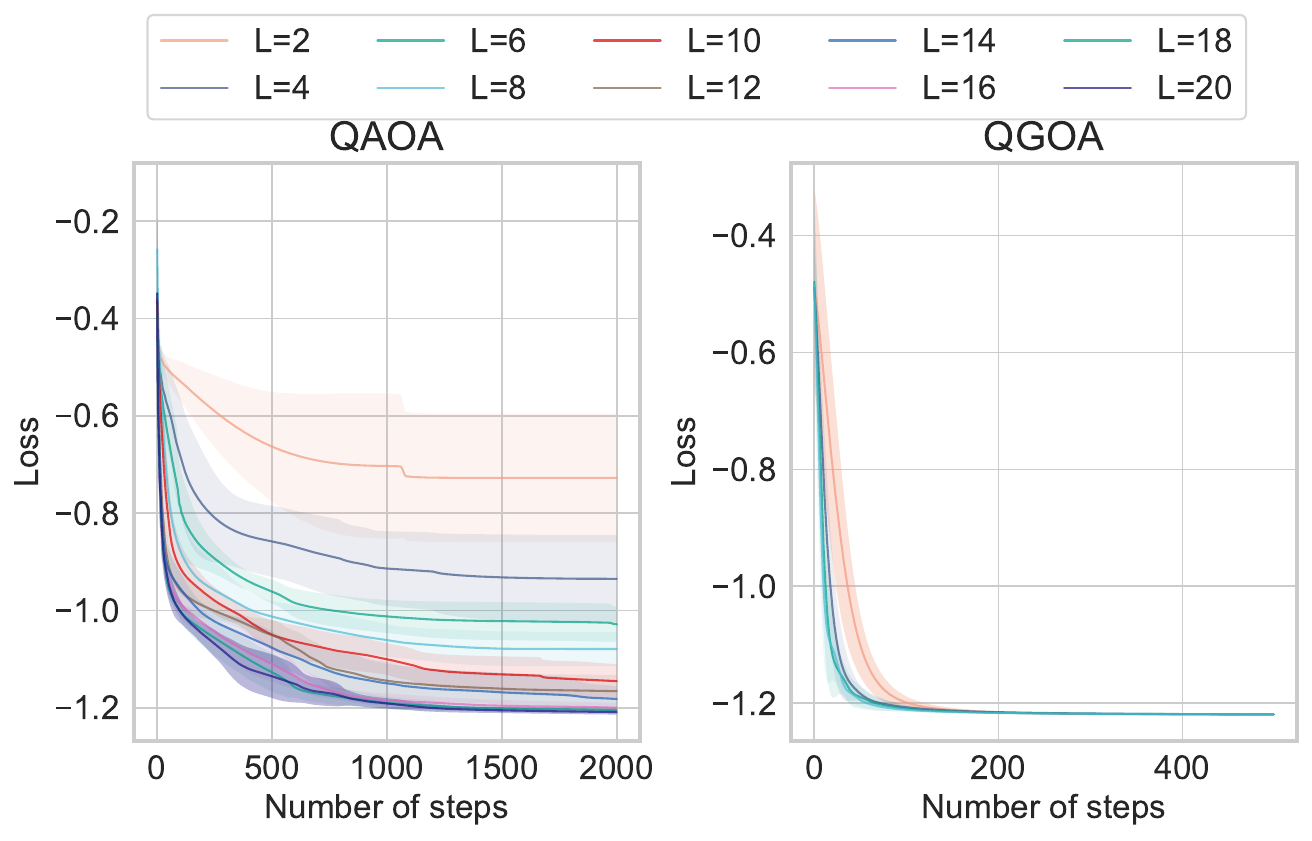}
    \end{minipage}\label{fig: finance_qaoa_QGCN}}
    \subfigure[]{
    \begin{minipage}[t]{0.42\textwidth}
    \centering
    \includegraphics[width=1\textwidth]{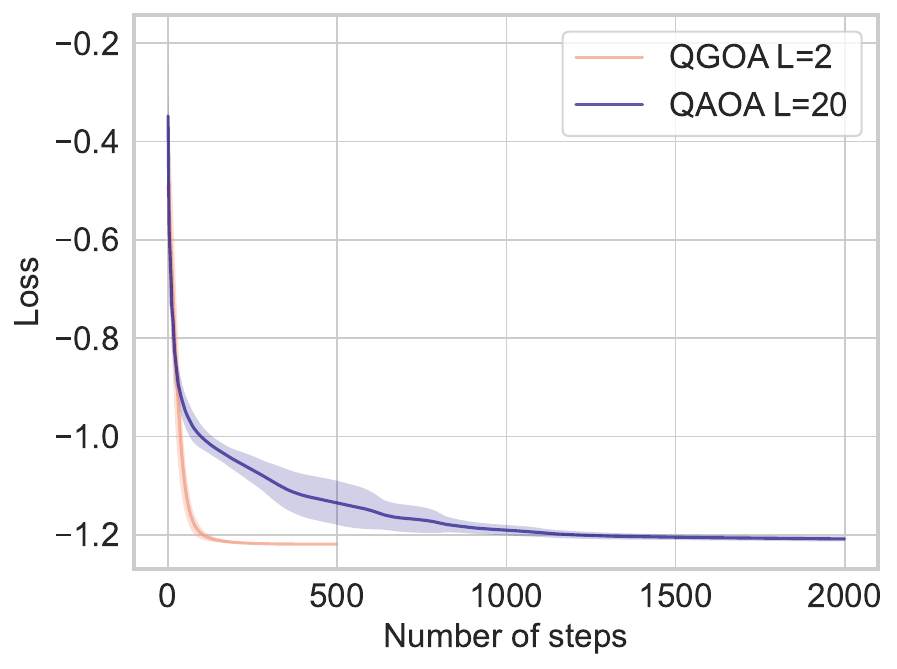}
    \end{minipage}\label{fig: finance_qaoa_qgcn_l}}
    \subfigure[]{
    \begin{minipage}[t]{0.35\textwidth}
    \centering
    \includegraphics[width=1\textwidth]{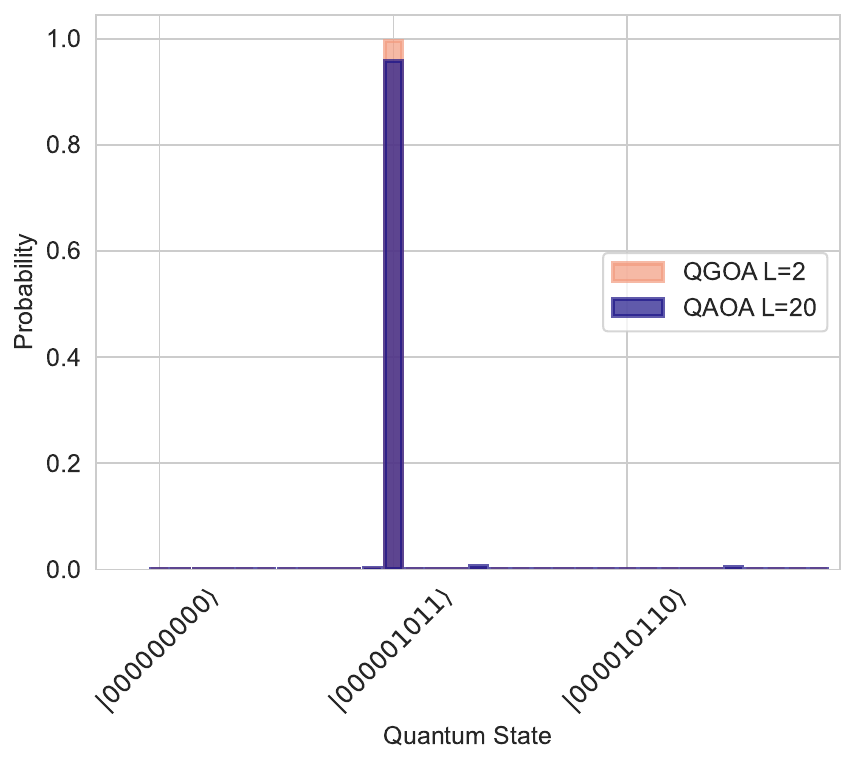}
    \end{minipage}\label{fig: finance_qaoa_qgcn_h}}
    \caption{The simulation results obtained using the QGOA and QAOA ansatz for the portfolio optimization problem are shown. (a) The construction of the QGOA and QAOA ansatz. (b) The performance of QGOA and QAOA with varying numbers of layers, ranging from 2 to 20. (c) A comparison between QGOA and QAOA with the best layout. (d) The probability distribution of the optimized assignments obtained by QGOA and QAOA.}
\end{figure*}

\subsubsection{Simulation Results} We compare our proposed method with the QAOA method, both of which are depicted in Fig.~\ref{fig: finance}, by conducting experiments on a nine-qubit portfolio optimization problem. We use 10 sets of random seeds and run both methods with different numbers of layers (2 to 20). While QAOA is trained for 2000 steps, QGOA is trained for only 500 steps. We assessed the performance of both methods based on their final objective function value and the probability distribution of the optimized assignment.

Fig.~\ref{fig: finance_qaoa_QGCN} displays the final objective function value, also known as the loss value, for different optimized layouts with 2 to 20 layers of QAOA and QGOA methods. It is evident that the QAOA method achieved the best result with an L=20 layout quantum circuit, while the QGOA method achieved the best result with an L=2 layout quantum circuit.

Fig.~\ref{fig: finance_qaoa_qgcn_l} displays a line chart comparing the performance of our proposed QGOA method and the QAOA method for solving the portfolio optimization problem. The y-axis represents the loss function given in Eq.~\ref{eq finance}, while the x-axis shows the number of iterations. The violet line represents the performance of the QAOA method, while the orange line represents the performance of our proposed QGOA method. The line chart shows that our proposed method has a slight advantage over the QAOA method. Furthermore, Fig.~\ref{fig: finance_qaoa_qgcn_h} shows a histogram representing the probability distribution of the optimized assignments obtained by each method. The violet and orange distributions correspond to the QAOA and QGOA methods, respectively. The exact solution to the problem is $``000001011"$, as obtained by brute force. The histogram reveals that our proposed method has a success rate of $99.9\%$ in finding the optimal solution, indicating that it is highly likely to find the exact solution. In contrast, the QAOA method has a success rate of $97\%$ in finding the optimal solution. These results demonstrate that our proposed QGOA method has a slightly higher performance than the QAOA method for solving the portfolio optimization problem.

We define $\text{N}_q$ as the number of qubits, $\text{L}$ as the number of layers, $\text{N}_e$ as the number of edges in the finance problem, $\text{N}_1$ as the number of single-qubit gates, $\text{N}_2$ as the number of two-qubit gates, $\text{T}$ as the number of iterations required for convergence, and $\text{N}_p$ as the number of parameters in each iteration. The quantum and classical costs of QAOA and QGOA are shown in Tab.~\ref{tab:finance_cost}. For QGOA, we define $\text{N}_1{=}(2*\text{N}_q+\text{N}_e)*\text{L}$, $\text{N}_2{=}2*\text{N}_e*\text{L}$, and $\text{N}_p{=}(2*\text{N}_q+1)*\text{L}$. For QAOA, we define $\text{N}_1{=}2*\text{N}_q*\text{L}+\text{N}_q$, $\text{N}_2{=}\text{N}_e*\text{L}$, and $\text{N}_p{=}2*\text{L}$. It is apparent that QGOA requires fewer single- and two-qubit gates and has a lower classical cost with $\text{T}*\text{N}_p$ classical resources.

\begin{table}[htbp]
    \centering
    \caption{The classical and quantum costs required for the nine-qubit portfolio optimization.}
    \begin{tabular}{lccccc}
         \toprule
         & &\multicolumn{2}{c}{Quantum-Cost} &\multicolumn{2}{c}{Classical-Cost} \\
         \cmidrule(lr){3-4}\cmidrule(lr){5-6}
         & $\text{L}$&$\text{N}_1$&$\text{N}_2$&$\text{T}$&$\text{N}_p$\\ 
         \midrule
         QGOA & 2 & 96 & 120  & 300 & 38\\
         QAOA & 20 & 369 & 600 & 1800 & 40\\
         \bottomrule
    \end{tabular}
    \label{tab:finance_cost}
\end{table}

\subsection{Minimum Vertex Cover Optimiztion}

\subsubsection{Minimum Vertex Cover Problem} The minimum vertex cover problem is a well-known combinatorial optimization problem in graph theory, which involves finding the smallest set of vertices in a graph such that every edge is incident to at least one vertex in the set. The minimum vertex cover problem can be formulated as a mathematical optimization problem with binary decision variables, where each variable represents whether or not a vertex is included in the cover. The objective is to minimize the number of vertices in the cover while ensuring that each edge in the graph is incident to at least one vertex in the cover, which can be written as:

\begin{align*}
\ell(x)=\arg \min_{\boldsymbol{x}} \quad & \sum_{(i,j)\in E} (1-x_i)(1-x_j)+b\sum_{i \in V} x_i.
\label{Eq_minivercover}
\end{align*}
where $\boldsymbol{x}$ is a vector of binary decision variables, with $x_i$ representing whether vertex $i$ is included or not included in the cover.

\subsubsection{QGOA for Minimum Vertex Cover Optimization} For the twelve-qubit minimum vertex cover problem, the graph quantum algorithm design consists of two blocks: two layers of single rotation gates and a layer of first-order $\text{H}_{T^{1st}}$-model. Next, the algorithm uses an observable M related to the problem. The objective function can be expanded as shown in Eq.~\ref{Eq_minivercover}. To embed this problem into the observable, we define M as:

\begin{align}
   M=\sum_{(i,j)\in E} (Z_i+Z_j+Z_i Z_j)/4 - \sum_{i \in V} Z_i/2,
\end{align}
where Z represents the Pauli-Z operator. After measuring using observable M, we obtain classical information, which is minimized by optimizing the parameters $\theta$ and $\eta$ of the graph quantum circuit using the Adam optimizer.

\begin{figure*}[htbp]
    \centering
    \subfigure[]{
    \begin{minipage}[t]{0.45\textwidth}
    \centering
    \includegraphics[width=1\textwidth]{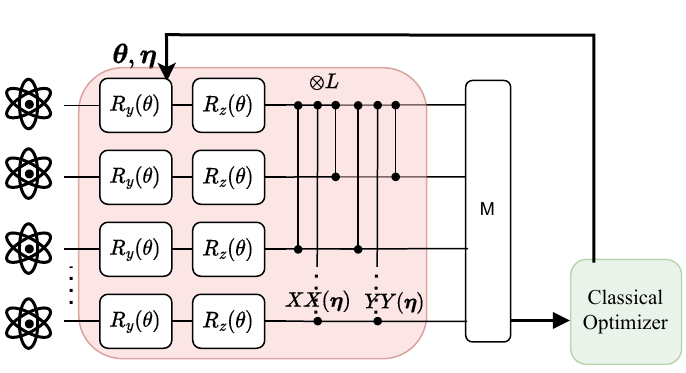}
    \end{minipage}\label{fig: graph_circuit}}
    \subfigure[]{
    \begin{minipage}[t]{0.45\textwidth}
    \centering
    \includegraphics[width=1\textwidth]{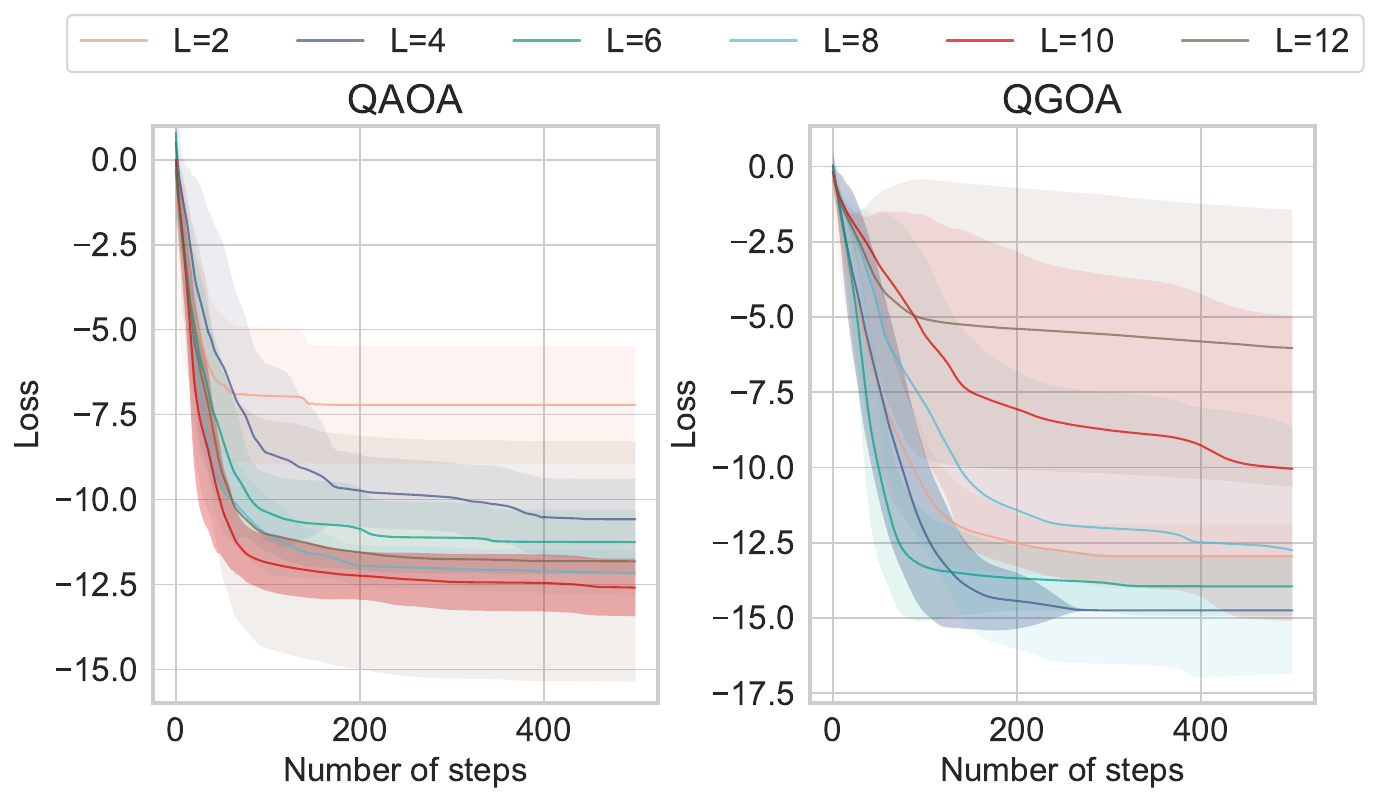}
    \end{minipage}\label{fig: graph_qaoa_qgcn}}
    \subfigure[]{
    \begin{minipage}[t]{0.36\textwidth}
    \centering
    \includegraphics[width=1\textwidth]{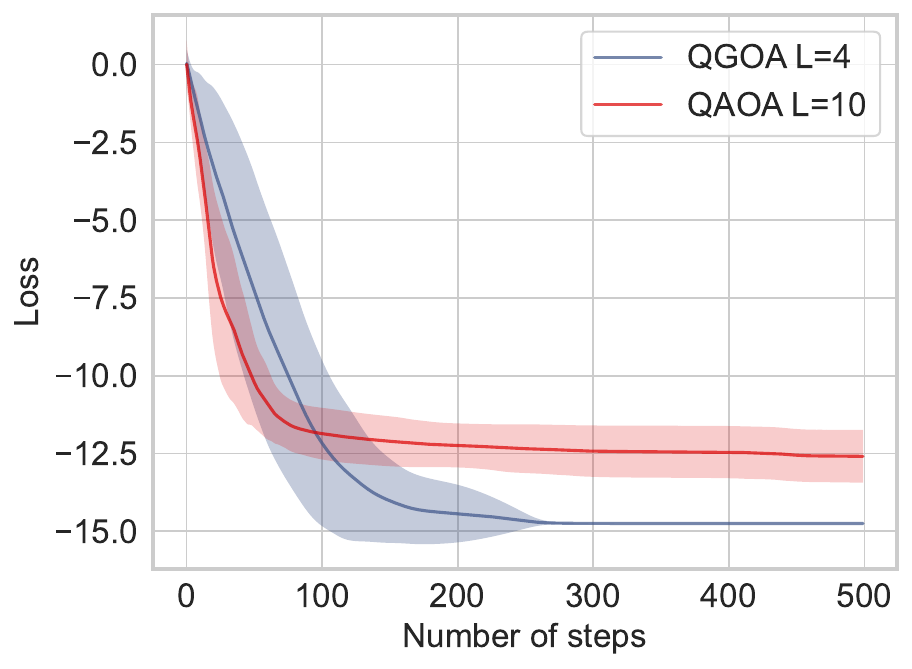}
    \end{minipage}\label{fig: graph_qaoa_qgcn_l}}
    \subfigure[]{
    \begin{minipage}[t]{0.30\textwidth}
    \centering
    \includegraphics[width=1\textwidth]{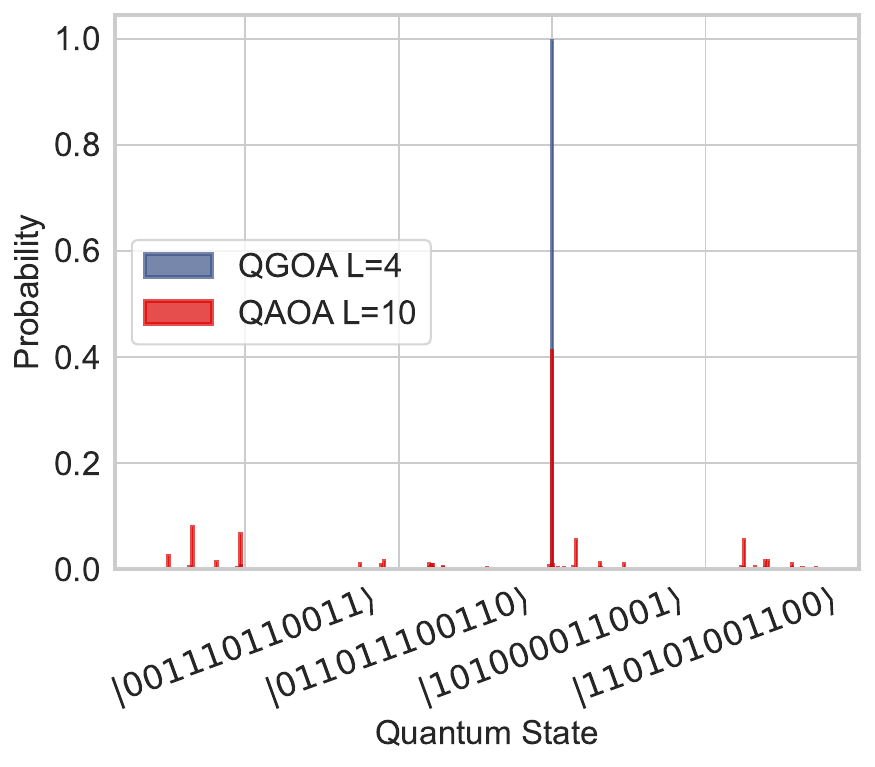}
    \end{minipage}\label{fig: graph_qaoa_qgcn_h}}
    \subfigure[]{
    \begin{minipage}[t]{0.20\textwidth}
    \centering
    \includegraphics[width=1\textwidth]{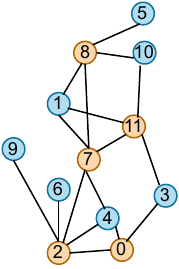}
    \end{minipage}\label{fig: graph}}
    \caption{The simulation results obtained using the QGOA and QAOA ansatz for the minimum vertex cover problem are shown. (a) The construction of the QGOA ansatz for minimum vertex cover problem. (b) The performance of QGOA and QAOA with varying numbers of layers, ranging from 2 to 12. (c) A comparison between QGOA and QAOA with the best layout. (d) The probability distribution of the optimized assignments obtained by QGOA and QAOA. (e) The final result of QGOA for the minimum vertex cover optimization.}
\end{figure*}

\subsubsection{Simulation Results} To compare the performance of our proposed method and the QAOA method, we conducted experiments on the twelve-qubit minimum vertex cover problem as shown in Fig.~\ref{fig: graph}. 

In order to determine the optimal layout for QAOA and QGOA methods, we evaluate the performance of optimized layouts with varying numbers of layers (2 to 12). For each algorithm, we choose six layouts and use 10 different sets of random seeds for each layout to ensure diverse initial parameters. We then execute all layouts for the same number of steps (500) and measure the final outcome to determine the best layout.

Fig.~\ref{fig: graph_qaoa_qgcn} displays the final loss value for different optimized layouts with 2 to 12 layers of QAOA and QGOA methods. It is evident that the QAOA method achieves the best result with an L=10 layout quantum circuit, while the QGOA method achieves the best result with an L=4 layout quantum circuit.

Fig.~\ref{fig: graph_qaoa_qgcn_l} displays a line chart that compares the performance of our proposed QGOA method and the QAOA method for solving the minimum vertex cover problem. The y-axis represents the loss function given in Eq.~\ref{Eq_minivercover}, while the x-axis shows the number of iterations. The red line represents the performance of the QAOA method, while the blue-purple line represents the performance of our proposed QGOA method. The line chart shows that our proposed method has a significant advantage over the QAOA method. Furthermore, Fig.~\ref{fig: graph_qaoa_qgcn_h} shows a histogram that represents the probability distribution of the optimized assignments obtained by each method. The red and blue-purple distributions correspond to the QAOA and QGOA methods, respectively. The success rate refers to the overlap between the maximally populated state of the optimized state and the exact solution. In this case, the exact result obtained by brute force is the bit string $``101000011001"$, which represents a cover consisting of the first, third, eighth, ninth, and twelfth vertices in a graph with twelve vertices. The histogram reveals that our proposed method has a success rate of $99.9\%$ in finding the optimal solution, indicating that it is highly likely to find the exact solution. In contrast, the QAOA method has a success rate of $42\%$ in finding the optimal solution. These results demonstrate that our proposed QGOA method has an apparently higher performance and faster convergence than the QAOA method for solving the minimum vertex cover problem.

We define $\text{N}_q$ as the number of qubits, $\text{L}$ as the number of layers, $\text{N}_e$ as the number of edges in the graph, $\text{N}_1$ as the number of single-qubit gates, $\text{N}_2$ as the number of two-qubit gates, $\text{T}$ as the number of iterations required for convergence, and $\text{N}_p$ as the number of parameters in each iteration. The quantum and classical costs of QAOA and QGOA are shown in Tab.~\ref{tab:finance_cost}. For QGOA, we define $\text{N}_1{=}2*\text{N}_q*\text{L}$, $\text{N}_2{=}2*\text{N}_e*\text{L}$, and $\text{N}_p{=}(2*\text{N}_q+1)*\text{L}$. For QAOA, we define $\text{N}_1{=}2*\text{N}_q*\text{L}+\text{N}_q$, $\text{N}_2{=}\text{N}_e*\text{L}$, and $\text{N}_p{=}2*\text{L}$. It is apparent that QGOA requires fewer single- and two-qubit quantum gates and has a higher classical cost with $\text{T}*\text{N}_p$ classical resources. Although QGOA requires higher classical resources, it achieves much better performance compared to QAOA. Moreover, in our experiments, increasing the classical resource requirements of QAOA does not improve its performance significantly as shown in Fig.~\ref{fig: graph_qaoa_qgcn}.

\begin{table}[htbp]
    \centering
    \caption{The classical and quantum costs required for the twelve-qubit minimum vertex cover problem.}
    \begin{tabular}{lccccc}
         \toprule
         & &\multicolumn{2}{c}{Quantum-Cost} &\multicolumn{2}{c}{Classical-Cost} \\
         \cmidrule(lr){3-4}\cmidrule(lr){5-6}
         & $\text{L}$&$\text{N}_1$&$\text{N}_2$&$\text{T}$&$\text{N}_p$\\ 
         \midrule
         QGOA & 4 & 96 & 136 & 250 & 100\\
         QAOA & 10 & 252 & 170 & 450 & 20\\
         \bottomrule
    \end{tabular}
    \label{tab:graph_cost}
\end{table}

\section{Scalability}
Tab.~\ref{tab:finance_cost} shows the classical and quantum resources utilized by QAOA and QGOA methods for portfolio optimization. As the system width increases, both quantum and classical resources increase for both methods. However, the QAOA model requires significantly more depth to converge compared to the QGOA model, as each layer has less freedom with only two parameters. The scalability of these advantages is crucial for NISQ quantum simulators, which have limited resources available.

We compared the resource consumption of QAOA- and QGOA-based quantum methods by considering the counts of quantum and classical resources. Specifically, the quantum resource count reflects the number of two-qubit quantum gates used in the computation, as the two-qubit gates are a major resource in quantum computing. On the other hand, the classical resources are determined by the product of the converged iteration count and the number of parameters. To simplify the analysis, we focused on sparse portfolio optimization cases, where the correlation between different assets in the portfolio is sparse. This means that only a few pairs of assets have a significant correlation, while most other pairs have little or no correlation. Sparse portfolio optimization is a common and challenging problem in practice, as it arises in various situations, such as investing in a diverse set of industries or regions, or alternative investments such as private equity or real estate. In these cases, the assets may have little correlation with each other due to their unique characteristics and market factors.

\begin{figure}[htp]
    \centering
    \includegraphics[width=1\columnwidth]{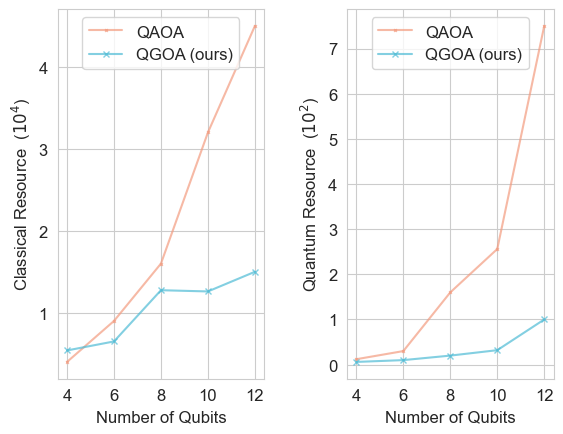}
    \caption{The required classical and quantum resources increase as the qubit system size grows for the portfolio optimization problem. The cost of QAOA is shown with orange lines, while the cost of QGOA is represented by blue lines.}
    \label{fig: scale}
\end{figure}

In the range of 4-qubits to 12-qubits, as shown in Fig.~\ref{fig: scale}, the QGOA method demonstrated significant advantages over the QAOA method in terms of both quantum and classical resources. These advantages become even more pronounced as the spatial dimension of the system increases. Moreover, the relative advantage of QGOA over QAOA is scalable, indicating that the QGOA method can allow us to tackle system sizes much larger than what the existing optimization method QAOA can handle, especially given the severe constraints of the NISQ era.

\section{Performance Evaluation of Various Ansatzes}
To evaluate the effectiveness of the QGOA method, we conduct several numerical experiments with different methods in nine qubits-portfolio optimization. These methods include some related works, such as QAOA~\cite{farhi2014quantum,mahroo2023learning}, XYQAOA~\cite{wang2020x}, BitflipQAOA~\cite{hadfield2019quantum}, , and EQC~\cite{skolik2023equivariant}. Compared to the optimal layouts of each algorithm presented in Fig.~\ref{fig: compare}, our algorithm not only demonstrates superior performance but also achieves the fastest coverage.
\begin{figure}[htp]
    \centering
    \includegraphics[width=0.8\columnwidth]{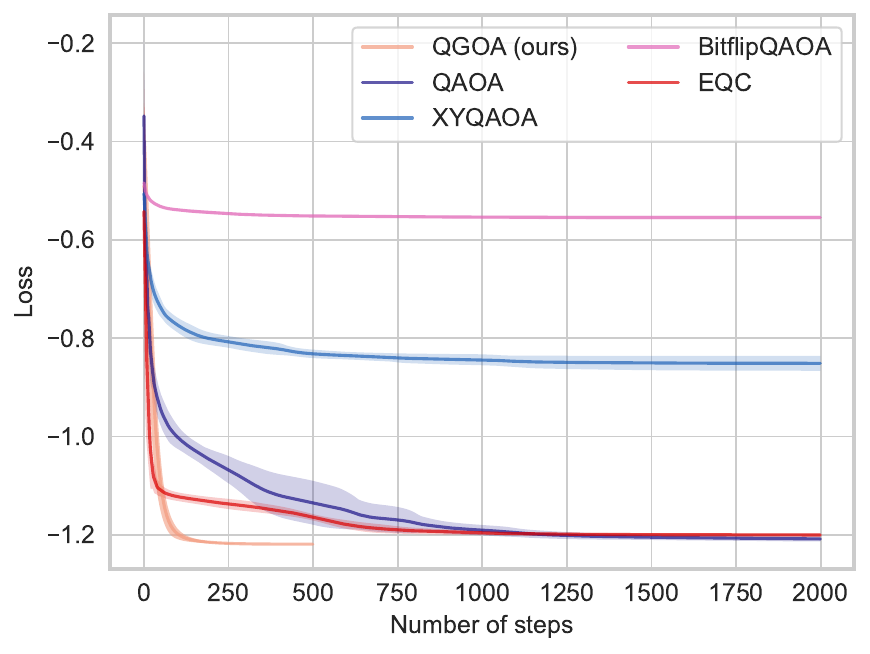}
    \caption{Comparison of simulation results: QGOA (our method) vs. QAOA, XYQAOA, BitflipQAOA, and EQC each using their optimal layouts. }
    \label{fig: compare}
\end{figure}

\section{Conclusion}

In the NISQ era, VQAs have shown great promise for solving problems in Euclidean space. However, extending these algorithms to non-Euclidean spaces, such as graph-related problems, presents a challenge. In this paper, we propose a quantum graph optimization algorithm that incorporates the principles of the message-passing mechanism that captures the geometric equivariance and symmetry present in graph-related problems.

To evaluate the effectiveness of our approach, we apply it to two different tasks: portfolio assignment, and minimum vertex cover. We compare our approach with commonly used graph algorithms, QAOA, XYQAOA, BitflipQAOA, and EQC to demonstrate the advantages of our method. Our results show that our algorithm has the potential to significantly improve the performance of graph-based approximate optimization tasks in quantum computing.

While our approach is still in its early stages, we believe that it represents an important step towards more efficient and effective quantum algorithms for graph-related problems. As quantum computing technology continues to advance, we expect that our approach will become increasingly important for solving a wide range of graph-based machine learning tasks.

\appendices
\section{Topological Graph Embedding in Quantum}\label{appendix:a}
In quantum computing, several efforts have been made to address graph problems using techniques such as QAOA~\cite{farhi2014quantum} and VQA~\cite{kandala2017hardware}. However, one limitation of VQE is that they do not explicitly encode the topological information of the problem. While QAOA encodes the logical information using the Ising model, it operates under the adiabatic approximation, which can require long-time evolution and may not be efficient for large-scale graphs. 

To overcome these limitations, our research was inspired by classical graph convolutional networks and aimed to explore topological graph embedding techniques in quantum computing. Specifically, we propose a method for encoding the topology of a graph into the Hamiltonian A. To encode the edge information, we designed a connection matrix as follows:
\begin{align}
A &{=} {\sum_{(i,j) \in \mathbb{E}}}w_{ij}(|1\rangle\langle 0|_i{\otimes} |0\rangle\langle 1|_j{+}|1\rangle\langle 0|_j{\otimes} |0\rangle\langle 1|_i){+}{\sum_{i \in \mathbb{V}}} w_{ii}|1\rangle\langle 1|_i\notag\\
&= \sum_{(i,j) \in \mathbb{E}} w_{ij}((\sigma_i^x-i\sigma_i^y)(\sigma_j^x+i\sigma_j^y)/4
\notag\\
&~~~~~~~~~~ +(\sigma_j^x-i\sigma_j^y)(\sigma_i^x+i\sigma_i^y)/4)
\notag\\
&~~~~~~~~~~ +\sum_{i \in \mathbb{V}} w_{ii}(1-\sigma_i^z)/2\notag\\
&= \sum_{(i,j) \in \mathbb{E}} w_{ij}(\sigma_i^x\sigma_j^x+i\sigma_i^x\sigma_j^y-i\sigma_j^x\sigma_i^y+\sigma_i^y\sigma_j^y\notag\\
&~~~~~~~~~~+\sigma_j^x\sigma_i^x+i\sigma_j^x\sigma_i^y-i\sigma_i^x\sigma_j^y+\sigma_j^y\sigma_i^y)/4\notag\\
&~~~~~~~~~~ +\sum_{i \in \mathbb{V}} w_{ii}(1-\sigma_i^z)/2\notag\\
&=\sum_{(i,j) \in \mathbb{E}} w_{ij}/2 (\sigma_i^x \sigma_j^x + \sigma_i^y \sigma_j^y)+\sum_{i \in \mathbb{V}} w_{ii}/2(1-\sigma_i^z)
\end{align}
where $w_{ij}$ is the wight of edge $E_{ij}$, and $|0\rangle\langle 1|=(\sigma^x+i\sigma^y)/2$, $|1\rangle\langle 0|=(\sigma^x-i\sigma^y)/2$, and $|1\rangle \langle 1|=(1-\sigma^z)/2$, and the $\sigma_i^x$, $\sigma_i^y$, $\sigma_i^z$ are the Pauli-X, Pauli-Y, Pauli-Z matrices acting on $i$-th qubit.

\section{Information Aggregation in Quantum}\label{app: quantum aggregation}
To perform a quantum aggregation operation, we first encode the feature information using the Ry gate. Next, we apply parameterized Ry and Rz gates. We then apply the first-order topological equivariance $\text{H}_{T^{1st}}$-model to the quantum state. This model captures the topological information of the graph. Finally, we use a measurement operator defined as:

\begin{align}
M = \begin{bmatrix}
1 & 0 \\
0 & -1
\end{bmatrix}
\end{align}

The combination of angle encoding using Ry gates, as well as parameterized Ry and Rz gates, along with the first-order topological equivariance $\text{H}_{T^{1st}}$-model and the measurement operator M, demonstrates a quantum aggregation operation.

To validate the aggregation operation, we assume the graph $G=(\mathcal{V}, \mathcal{E})$ with 3 nodes $V_1, V_2, V_3$ and $\mathcal{E}={e_{1,3}, e_{1,2}}$. We encode the topology of the graph into the quantum system. We suppose the default initial quantum state is $|0\rangle$, then apply the angle encoding quantum gate $R_y(x)$, then apply  the parameterized quantum gate $R_y(\theta)$, $R_z(\theta)$, and the first order $\text{H}_{T^{1st}}$-model on $e_{1,3}, e_{1,2}$ with parameter $\eta$. When we measure the first qubit by M, we obtain the $\langle M_1 \rangle$. When we measure the second qubit by M, the final result is $\langle M_2\rangle$. When we measure the third qubit by M, the final result is $\langle M_3\rangle$.  $F_i({\theta}_i)$ means the combination of trigonometric functions with ${\theta}_i$. $X_i$ to represent a term consisting of a combination of $i$ trigonometric functions.

\begin{align}
\langle M_1\rangle&=  \sum F_{i}({\theta}_i) X_{1i}(x_1)+\sum F_{i}({\theta}_i) X_{1i}(x_2)\notag\\
&+\sum F_{i}({\theta}_i) X_{1i}(x_3)+\sum F_{i}({\theta}_i) X_{2i}({x}_i)\notag\\
&+\sum F_{i}({\theta}_i) X_{3i}({x}_i)
\end{align}

\begin{align}
\langle M_2\rangle&= \sum F_{i}({\theta}_i) X_{1i}(x_1)+\sum F_{i}({\theta}_i) X_{1i}(x_2)\notag\\
&+\sum F_{i}({\theta}_i) X_{2i}({x}_i)+\sum F_{i}({\theta}_i) X_{3i}({x}_i)
\end{align}

\begin{align}
\langle M_3\rangle&= \sum F_{i}({\theta}_i) X_{1i}(x_1)+\sum F_{i}({\theta}_i) X_{1i}(x_3)\notag\\
&+\sum F_{i}({\theta}_i) X_{2i}({x}_i)+\sum F_{i}({\theta}_i) X_{3i}({x}_i)
\end{align}

The results show that nearby node information from nodes that are connected by an edge is directly aggregated. However, information from non-neighboring nodes, which are not directly connected by an edge, is aggregated indirectly. The two-parameter and three-parameter terms represent this in the results.  

This demonstrates a form of aggregation in quantum algorithm graph design. We aggregate topological information. Nearby information is directly aggregated, while global information appears in the two- and three-parameterized terms, indicating information is non-directly aggregated in our algorithm.

\section*{Acknowledgment}
The authors of this work would like to extend their sincere gratitude to HKUST and HSBC for their invaluable support and contribution to this research project. Additionally, we are deeply grateful to all the individuals from both parties who generously contributed their time and expertise to this work. Their contributions were essential to the success of the project, and we are truly thankful for their dedication and effort.

\bibliographystyle{IEEEtran}

\end{document}